\documentclass[%
 aip,
 amsmath,amssymb,
]{revtex4-1}

\usepackage{graphicx}
\usepackage{dcolumn}
\usepackage{bm}

\usepackage[utf8]{inputenc}
\usepackage[T1]{fontenc} 
\usepackage[normalem]{ulem}
\usepackage[cmtip,all]{xy} 
\usepackage{comment}
\usepackage{multirow}
\usepackage{mathtools}
\usepackage{soul}
\usepackage{latexsym}
\usepackage{amsmath}
\usepackage{mathrsfs}
\usepackage{bm}
\usepackage{amsthm}
\usepackage{physics}
\usepackage{bbm}
\usepackage{amsfonts}
\usepackage{amssymb}
\usepackage[usenames]{color}
\usepackage{epsfig}
\usepackage{hyperref}
\usepackage{chemformula}

\begin{document}

\preprint{AIP/123-QED}

\title[]{Intrinsic Rashba coupling due to Hydrogen bonding in DNA}

\author{S.Varela}
\email{svarela@yachaytech.edu.ec}
\affiliation{Yachay Tech University, School of Chemical Sciences \& Engineering, 100119-Urcuqu\'i, Ecuador}

\author{B. Monta\~nes}
\affiliation{Laboratorio de F\'isica Estad\'istica de Sistemas Desordenados, Centro de F\'isica, Instituto 
Venezolano de Investigaciones C\'ientificas (IVIC), Apartado 21827, Caracas 1020 A, Venezuela}

\author{F. L\'opez}
\affiliation{Yachay Tech University, School of Chemical Sciences \& Engineering, 100119-Urcuqu\'i, Ecuador}

\author{B. Berche}
\altaffiliation[Also at ]{Yachay Tech University, School of Physical Sciences \& Nanotechnology, 100119-Urcuqu\'i, Ecuador}
\affiliation{Laboratoire de Physique et Chimie Th\'eoriques, UMR Universit\'e de Lorraine-CNRS 7019 54506 Vand\oe uvre les Nancy, France}

\author{B. Guillot}
\affiliation{Universite de Lorraine, Institut Jean Barriol, Laboratoire de Cristallographie, R\'esonance Magn\'etique et Mod\'elisations 
CRM2, UMR CNRS-UL 7036, France}

\author{V. Mujica}
\affiliation{School of Molecular Sciences, Arizona State University, Tempe, Arizona 85287-1604, USA}

\author{E. Medina}
\altaffiliation[Also at ]{Laboratorio de F\'isica Estad\'istica de Sistemas Desordenados, Centro de F\'isica, Instituto 
Venezolano de Investigaciones C\'ientificas (IVIC), Apartado 21827, Caracas 1020 A, Venezuela}
\email{emedina@yachaytech.edu.ec}
\affiliation{Yachay Tech University, School of Physical Sciences \& Nanotechnology, 100119-Urcuqu\'i, Ecuador}

\date{\today}

\begin{abstract}
We present an analytical model for the role of hydrogen bonding on the spin-orbit coupling of model DNA molecule. Here we analyze in detail the electric fields due to the polarization of the Hydrogen bond   on the DNA base pairs and derive, within tight binding analytical band folding approach, an intrinsic Rashba coupling which should dictate the order of the spin active effects in the Chiral-Induced Spin Selectivity (CISS) effect. The coupling found is ten times larger than the intrinsic coupling estimated previously and points to the predominant role of hydrogen bonding in addition to chirality in the case of biological molecules. We expect similar dominant effects in oligopeptides, where the chiral structure is supported by hydrogen-bonding and bears on orbital carrying transport electrons.

\end{abstract}

\maketitle

The Chiral-Induced Spin Selectivity (CISS) effect is a surprisingly strong spin polarization effect induced by chiral molecular structures (either point or \mbox{globally} chiral) in the absence of magnetic centers and for \mbox{relatively} light atoms such as carbon and nitrogen\cite{RayNaaman,NaamanSingleMolec,Gohler}. It was first proposed that the spin active ingredient to the observed electron spin polarization was the Spin-Orbit coupling (SO) \cite{Sina} (preserving time reversal symmetry) because the chiral molecules, i.e., Amino acids, Oligopeptides, DNA, etc. do not sustain spin order of any kind. The source of the SO interaction was proposed to be of atomic origin (as in BIA semiconductors) \mbox{theoretically} predicting polarizations of up to a few percent in second order Born scattering\cite{MedinaLopez} and bearing all the signatures (except for magnitude) found in asymptotically free electrons experiments\cite{Gohler}. Furthermore, bound electron transport single molecule experiments\cite{NaamanSingleMolec} yields much larger polarizations (+50\%) for just a few tens of turns of a DNA helix. The polarizing power is so significant that such a setup has been successfully used to produce magnetic memories without permanent magnets\cite{PaltielMemory}.

Recent theoretical proposals to explain single molecule experiments suggest the existence of strong internal electric fields, other than those from the atomic cores, that yield Rashba type interactions\cite{Cuniberti,GutierrezAdame,GuoSun}. Since all such approaches incorporate time reversal symmetry (exception is ref.\cite{GuoSun}), only broken by choice of transport direction, they yield very similar spin filtering scenarios. Nevertheless, the sources of the specific fields involved has not been identified. In this work, we analyze the electric field due to the polarized nature of the Hydrogen bond coupling the bases together in DNA. We find such fields yield a particularly strong Stark interaction on the $\pi$ orbital of carbon and oxygen sites which combined with their atomic spin-orbit interaction, yield an intrinsic Rashba effect. These Hydrogen bonds, in the case of DNA, are strengthened and protected from solvent hydration\cite{DNAHydration} and screened by the hydrophobic stacking of the bases which is a major contributor to the double helix stability.

 Non-local DFT calculations, modeling electric field profiles with intrabond resolution, have been recently performed \cite{Blanco-Ruiz}.
 Such computations bear on the near field electric potentials that are relevant to this study and have been known for some time\cite{ReviewHol}. In the case of the A-T base pair, the electric field generates a Stark interaction on the oxygen, double bonded to a carbon, on the thymine side. The second Hydrogen bond generates a similar Stark interaction on the N atom double bonded to a C on the Adenine base. The G-C base pair is asymmetrical yielding only a strong Stark interaction on a double bonded oxygen on the Guanine base, while on the Cytosine base two atoms, N and O (both double bonded), see a strong electric field. 
 
 Here we focus on the double bonded atoms because they provide the most mobile electrons ($\pi$ electrons) for inter-base pair processes\cite{Simserides,Varela}. We study the potential gradient lines derived from the electronic density of the bonded structures. The electric field profiles are in excellent agreement with those derived from DFT\cite{Blanco-Ruiz}. For H-X distances of 1.5 \AA, an intrabond electric field of 15 V/\AA~ and 35 V/\AA ~for can be found for the adenine and thymine bases respectively. Similar fields are produced on the identified site for the Guanine and Cytosine base pairs. These values of electric fields are of the same order as those seen in the hydrogenic atom at two Bohr radii from the nucleus (38 V/\AA) and stronger than any other sources of electric fields we have found in the DNA molecule. 

In this work we will derive, in perturbation theory, using the Slater-Koster tight-binding approach, an intrinsic Rashba coupling we believe is quantitatively responsible for the strong spin activity observed experimentally. The explicit analytical form for the spin-coupling derived has as source the atomic spin-orbit coupling of the carbon present on the DNA bases and the Stark interaction coupling the $\pi$ and $\sigma$ structures. The approach here has been well tested in low dimensional systems starting with graphene itself\cite{Huertas,Konschuh}, bilayer graphene\cite{McCann}, proximity effects to metallic surfaces both of noble\cite{Marchenko,Lopez}, and ferromagnetic metals\cite{McDonald,Peralta}, and coarse grained spin active models for DNA\cite{Varela}.

\section{DNA Model and Hydrogen bonding}

A detailed tight-binding model of DNA considering a single $\pi$ orbital per base and coupled to neighbouring bases on the same  and between helices has been proposed by Varela et al\cite{Varela2016}. The spin activity in the absence of magnetic centers and external magnetic fields is derived from the atomic spin-orbit interaction in the range of that of nitrogen, oxygen and carbon in the meV range. The model includes a full description of the helical geometry disposition of the orbitals so as to obtain explicit forms for the parameters of the kinetic and a number of intrinsic and Rashba spin-orbit couplings. The latter interaction was derived from an externally applied electric field on the DNA axis. The perturbative treatment is shown to preserve time reversal symmetry expected of the microscopic interactions i.e. the atomic spin-orbit and the Stark coupling.

In the absence of interactions in a tight-binding model, hydrogenic atomic orbitals are orthogonal on-site and the neighboring orbitals are coupled by wave-functions overlaps\cite{SlaterKoster}. When the spin orbit interaction associated to electric fields (internal and external to the molecule) are considered, new couplings appear between local orbitals and electron transfer paths are opened between electron bearing orbitals that can be spin-active. 
\begin{figure}[h]
\centering
	\includegraphics[ width=10cm]{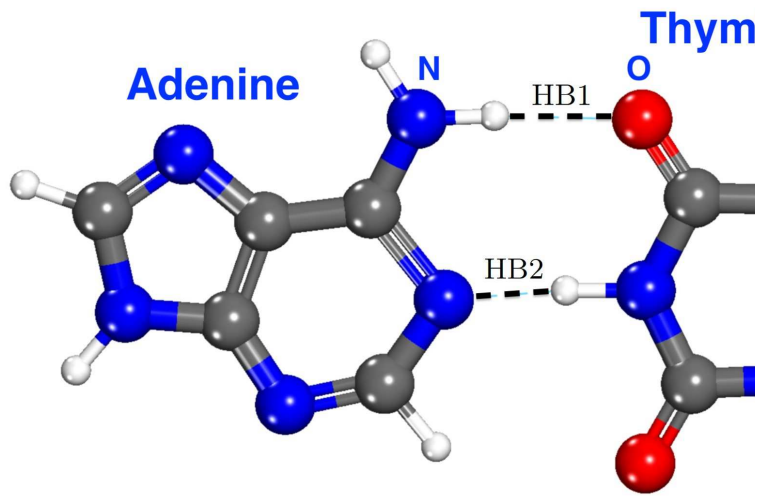}
	\caption{Adenine-Thymine base pair joined by two oppositely oriented Hydrogen bonds. The Hydrogen bond labelled HB1 is attached to the double bonded Oxygen atom (red) on the Thymine, while the HB2 bond is attached to the Nitrogen bond (blue) on the Adenine. Such double bonded atoms are assumed to provide the transport electrons on the structure.}
	  \label{etiquetas1}
\end{figure}
In the model of reference \cite{Varela2016}, only on site spin-orbit coupling and an externally applied electric field (along DNA axis) was considered. The resulting spin activity was determined to be in the neV to meV range for reasonable values of externally applied electric fields. Here, we consider internal sources of electric fields that have been ignored, but that are both much larger than can be externally applied and are also ubiquitous in biological molecules that have been tested for spin activity, i.e., hydrogen-bonding. 

In Fig.\ref{etiquetas1} we depict a adenine-thymine base pair. In direct contact with the Hydrogen bond, there is a double bonded oxygen is on the thymine side, while on the left hand side there is a double bonded nitrogen atom. We consider the $\pi$ orbital as a source of hopping electrons between bases\cite{Varela2016} subject to the hydrogen-bond effect. Other $\pi$ orbitals on each of the bases will provide electrons but their spin-activity will be shown to be at least three orders of magnitude smaller. We thus single out the doubly bonded oxygen on the thymine and the doubly bonded nitrogen on the adenine as part of the transport model. Later, we will show that this makes for a very testable model regarding mechanical deformations of the molecule leading to experimental predictions.

For each target atom, in the vicinity of the hydrogen bond, we consider an electric field of the form ${\bf
E}=E_x\hat x+E_y\hat y+E_z\hat{z}$, where $\hat{z}$ is the direction of the DNA axis, and $\hat{x}$ points along the Hydrogen bond. This system of coordinates rotates with the double helix. The electric field from the Hydrogen bond is considered to couple orbitals of the hydrogenic basis describing the double bonded atoms in its vicinity. Thus, it will induce a Stark interaction that couples the atomic orbitals of the appropriate symmetry. The Hamiltonian associated to this field can be written in spherical coordinates in the rotating basis as
\begin{equation}
    {H_s}=-eE_x r \sin{\theta}\cos{\phi} -eE_y r \sin{\theta}\sin{\phi} -eE_z r\cos{\theta},
\end{equation}
where $e$ is the electric charge. The orbitals associated with the local basis in spherical coordinates are
\begin{eqnarray}
   \psi_s({\bf r})=\langle{\bf r}\ket{s} &=& \frac{Z^3}{\sqrt{8\pi a_0^{3}}}e^{-Zr/2a_0}\left(1-\frac{Zr}{2a_0}\right),\nonumber\\
  \psi_x({\bf r})=\langle{\bf r} \ket{p_x} &=& \frac{Z^5}{\sqrt{32\pi a_0^{3}}}e^{-Zr/2a_0}\frac{Zr}{2a_0}\sin{\theta}\cos{\phi},\nonumber\\
   \psi_y({\bf r})=\langle{\bf r}  \ket{p_y} &=& -\frac{Z^5}{\sqrt{32\pi a_0^{3}}}e^{-Zr/2a_0}\frac{Zr}{2a_0}\sin{\theta}\sin{\phi},\nonumber\\
   \psi_z({\bf r})=\langle{\bf r} \ket{p_z} &=& \frac{Z^5}{\sqrt{32\pi a_0^{3}}}e^{-Zr/2a_0}\frac{Zr}{2a_0}\cos{\theta},
\end{eqnarray}
    with $Z$ the atomic number. 
    
The only non-zero elements coupled by the perturbing $H_s$ are
\begin{equation}
  \bra{2p_{x,y,z}}H_s\ket{2s}=\xi_{sp}^{x,y,z}.
\end{equation}

The effective coupling between $p_z$ orbitals ($\pi$ orbitals we have identified) on different base pairs, is represented by paths that involve the Stark and SO interactions, and the Slater-Koster overlaps, $E_{\mu\mu'}^{\imath\jmath}$, that connect $\mu$ orbital in site $\imath$ with $\mu'$ orbital in site $\jmath$. To the first order in perturbation theory, the $p_z-p_z$ coupling between two sites $\imath$ and $\jmath$, the paths for the Rashba interaction are for the interactions considered\cite{Varela2016, PastawskiMedina}

\begin{equation}
\underbrace{
 p_z^{\imath}\rightarrow { \xi_{p} }\rightarrow p_{x,y}^{\imath}\rightarrow {\xi_{sp}^{x,y}(\imath)}
}_{\hbox{onsite}}
\rightarrow
\underbrace{s^{\imath}  \rightarrow {E_{sz}^{\imath\jmath}} \rightarrow p_z^{\jmath}
}_{\hbox{overlap}},
 \label{Rplane}
\end{equation}

\begin{figure}[h]
\centering
	\includegraphics[width=10cm]{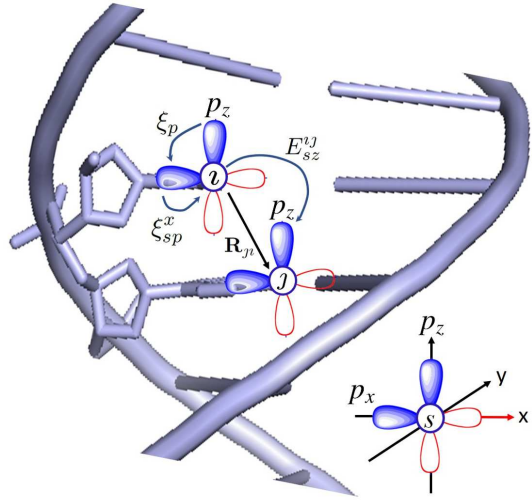}
	\caption{Rashba coupling to lowest order in perturbation theory due to the in plane (x-y plane on each base) electric field produced by the hydrogen bond. The figure depicts the orbital basis considered, and overlaps, associated with double bonded N or O under the effect of Hydrogen bond polarization.}
	  \label{paths}
\end{figure}

\newcommand{\longsquiggly}{\xymatrix{{}\ar@{~>}[r]&{}}}

\begin{equation}
\textcircled{$\imath$}\ \underbrace{
 p_z^{\imath}\stackrel{ { \xi_{p} }}{\longrightarrow} p_{x,y}^{\imath} \stackrel{\xi_{sp}^{x,y}(\imath)}{\longsquiggly}
s^\imath}_{\hbox{onsite}}\ \textcircled{$\imath$}\ 
\underbrace{s^{\imath}  \stackrel{{E_{sz_{\vphantom X}}^{\imath\jmath}}} {\Longrightarrow} p_z^{\jmath}
}_{\hbox{overlap}}\ \textcircled{$\jmath$},
 \label{Rplane}
\end{equation}

\begin{equation}
\underbrace{
 p_z^{\imath}\rightarrow { \xi_{p} }\rightarrow p_{x,y}^{\imath}
}_{\hbox{onsite}}
\underbrace{
\rightarrow {E_{(x,y),s}^{\imath\jmath}} \rightarrow
}_{\hbox{overlap}}
\underbrace{
s^{\jmath}  \rightarrow {\xi_{sp}^{z}(\jmath)}\rightarrow p_z^{\jmath}
}_{\hbox{onsite}},
 \label{Raxis}
\end{equation}

\begin{equation}
\textcircled{$\imath$}\ \underbrace{
 p_z^{\imath}\stackrel { \xi_{p} }{\longrightarrow} p_{x,y}^{\imath}
}_{\hbox{onsite}}
\ \textcircled{$\imath$}\ 
\underbrace{
p_{x,y}^{\imath}
\xRightarrow{E_{(x,y),s_{\vphantom X}}^{\imath\jmath}}
s^{\jmath}}_{\hbox{overlap}}\ \textcircled{$\jmath$}\ 
\underbrace{
s^{\jmath}  \stackrel {\xi_{sp}^{z}(\jmath)}{\longsquiggly} p_z^{\jmath}
}_{\hbox{onsite}}\ \textcircled{$\jmath$},
 \label{Raxis}
\end{equation}
where $\xi_p=\lambda\hbar^2/2$ represents the magnitude of the SO atomic interaction. For oxygen and nitrogen atoms, for example, $\xi_p\sim 9$ and $10$ meV respectively.
Paths in expression (\ref{Rplane}) are related to local electric field in the plane (see Fig. \ref{paths}), and the paths in (\ref{Raxis}) with the local electric field in helix-axis direction. 
  
\section{Rashba Coupling: band folding}
We can describe the coupling between different base pairs on the DNA double helix by considering a subspace $h_{\theta}$ given by the unperturbed $\pi$ orbitals at nearest neighbour bases $\imath-\jmath$ and their overlaps $E_{zz}^{\jmath\imath}$ as
\begin{equation}
h_{\theta}=
\left(\begin{array}{cc}
  \epsilon_{2p}^{\pi} & E_{zz}^{\imath\jmath} \\
  E_{zz}^{\jmath\imath}   & \epsilon_{2p}^{\pi}
\end{array}\right),
\label{htheta}
\end{equation}
where $\epsilon_{2p}^{\pi}$ are the unpertubed orbital energies. The SO and Stark interactions will couple the $\pi$ electrons to the on-site $2s$ and $2p$ orbitals described by the subspace  
\begin{equation}
h_{\chi}=\left(
\begin{array}{cccccc}
 \epsilon_s & \xi_{sp}^{x}(\imath)  & \xi_{sp}^{y}(\imath) & 0 & E_{sx}^{\imath\jmath} & E_{sy}^{\imath\jmath}  \\
   \xi_{sp}^{x}(\imath) &  
  \epsilon_{2p}^{\sigma} & 0  & E_{xs}^{\imath\jmath} & 0 & 0 \\
   \xi_{sp}^y(\imath) & 0 & \epsilon_{2p}^{\sigma} & E_{ys}^{\imath\jmath} & 0 & 0 \\
   0 & E_{sx}^{\jmath\imath} & E_{sy}^{\jmath\imath} &   \epsilon_s & \xi_{sp}^{x}(\jmath) & \xi_{sp}^{y}(\jmath) \\ 
     E_{xs}^{\jmath\imath} & 0 & 0 & \xi_{sp}^{x}(\jmath) &  \epsilon_{2p}^{\sigma}  & 0\\
     E_{ys}^{\jmath\imath} & 0 & 0 & \xi_{sp}^{y}(\jmath) & 0& \epsilon_{2p}^{\sigma}  \\
 \end{array}
\right). 
\label{hxi}
\end{equation}
Both sectors will be connected by the interactions and overlaps that connects all orbitals of two sites ($\imath$ and $\jmath$) in the helix in the form shown in (\ref{Rplane}) and (\ref{Raxis})
\begin{equation}
   H= \left(\begin{array}{cc}
         h_{\theta}& u   \\
        u^{\dagger} & h_{\chi} 
    \end{array}\right),
\end{equation}
where the matrix $u$ is given by
\begin{equation}
 u=\left(
 \begin{array}{cccccc}
    \xi_{sp}^z(\imath)  & -is_y\xi_p & is_x\xi_p & E_{zs}^{\imath\jmath} & 0 & 0 \\
    E_{zs}^{\jmath\imath} &0 &0  &\xi_{sp}^{z}(\jmath) & -is_y\xi_p &  is_x\xi_p
 \end{array}
 \right),
 \label{usubspace}
\end{equation}

The $u$ subspace contains the overlaps between the orbitals $p_z$ with the orbitals $s$, $p_z$ and $p_{x,y}$. Finally, the sub-space $h_{\chi}$ contains in the diagonal the energies $\epsilon_s$, $\epsilon_{2p}^{\pi}$ and $\epsilon_{2p}^{\sigma}$ , of the coupled orbitals $s$, $p_{x}$, $p_{y}$ respectively, and the off-diagonal the coupling between these orbitals. We assume that orbitals $p_x$ and $p_y$ are sigma bonded in the plane of the DNA helix and have the same energies and contrasts with the energy with the $\pi$ orbital at the same site.  

The eigenvalue equation to solve is 
 \begin{equation}
     \left( \begin{array}{cc}
       h_{\theta}   & u \\
        u^{\dagger}  & h_{\chi} 
     \end{array}
     \right)\left(\begin{array}{c}
         \theta   \\
           \chi
     \end{array}\right)=E\left(\begin{array}{c}
         \theta   \\
           \chi 
     \end{array}\right),
 \end{equation}
and the wave functions $\theta$ and $\chi$ are coupled by the $u$ sub-space. 
 Solving to eliminate the wave function subspace $\chi$, and taking linear order in $E$ and in the interactions, one obtains
\begin{equation}
    S^{-1/2}\left[H_{\theta} -u\left(H_{\chi}\right)^{-1}u^{\dagger} \right]S^{-1/2}\Phi \approx  E\Phi, 
\end{equation}
where $S=1+u(h_{\chi})^{-2}u^{\dagger}$, and we have defined $\Phi=S^{1/2}\theta$ as a normalized function to the same order as the effective Hamiltonian. We approximate $S\sim \mathbb{I}$ since no changes are brought to this order from the $u(h_{\chi})^{-2}u^{\dagger}$ correction. The effective Hamiltonian that couples $p_z-p_z$ orbitals, to the first order, is 
\begin{equation}
H_{eff} =    H_{\theta} -u\left(H_{\chi}\right)^{-1}u^{\dagger}.
\label{EffectiveH}
\end{equation}
The non-diagonal elements in $H_{eff}$ represent the Rashba interaction connecting $\imath$ and $\jmath$ sites. Substituting matrices (\ref{htheta}), (\ref{usubspace}) and (\ref{hxi}) in equation (\ref{EffectiveH}) and taking the reference Fermi level equal to the energy of the orbital $p_z$, that is to say, equal to $\epsilon_{2p}^{\pi}$, the effective coupling Rashba, $H_{R}$, is

\begin{equation}
    H_R=i\sum_{\imath\jmath}c_{\imath}^{\dagger}\left(\lambda_R^{x}s_y+\lambda_R^{y}s_x  + \lambda_R^{z}s_x \right)c_{\jmath},
\end{equation}
where 
\begin{equation}
\lambda_R^{x}=-\frac{\xi_p E_{sz}^{\imath\jmath}\left[\xi_{sp}^{x}(\imath)+\xi_{sp}^{x}(\jmath)\right]}{\left(\epsilon_{2p}^{\pi}-\epsilon_s\right)\left(\epsilon_{2p}^{\pi}-\epsilon_{2p}^{\sigma}\right)},
\end{equation}
\begin{equation}
\lambda_R^{y}=\frac{\xi_p E_{sz}^{\imath\jmath}\left[\xi_{sp}^{y}(\imath)+\xi_{sp}^{y}(\jmath)\right]}{\left(\epsilon_{2p}^{\pi}-\epsilon_s\right)\left(\epsilon_{2p}^{\pi}-\epsilon_{2p}^{\sigma}\right)},
\end{equation}
are the magnitudes of the Rashba interactions related to the electric fields in the plane, and 
\begin{equation}
\lambda_R^z=\frac{\xi_p E_{sy}^{\imath\jmath}\left[\xi_{sp}^{z}(\imath)+\xi_{sp}^{z}(\jmath)\right]}{\left(\epsilon_{2p}^{\pi}-\epsilon_s\right)\left(\epsilon_{2p}^{\pi}-\epsilon_{2p}^{\sigma}\right)},
\end{equation}
is the magnitude related with the electric field on the helix axis. $\imath$ and $\jmath$ labels in $\xi_{sp}$ indicate the component of the electric field. Perturbation theory, in this case, assumes that $\epsilon_{2p}^{\pi}$ and $\epsilon_{2p}^{\sigma}$ are non degenerate. The bare energies in perturbation theory are $\epsilon_s=-17.52$ eV, $\epsilon_{2p}^{\pi}=-8.97$ eV. The energy shift of the $\epsilon_{2p}$ due to inplane $\sigma$ bonding $\epsilon_{2p}^{\sigma}$ can be estimated by simple extended-Huckel theory 
to yield $\epsilon_{2p}^{\pi}-\epsilon_{2p}^{\sigma}\approx 2.5$ eV \cite{Harrison}.  All Rashba SO couplings are bi-linear in the atomic SO interaction and the Stark interaction. The helical geometry is directly involved in the order of the coupling if the external Stark field is in the $z$ direction. Nevertheless, this is not the case for the Hydrogen-bond source field, which is weakly dependent on the pitch. This fact is a testable result for this model. Note we ignore other sources of the Rashba coupling due to other external or internal electric fields. The contributions from fields on the axis were addressed in detail in ref.\cite{Varela}.

\section{Hydrogen bond and effective Rashba coupling strength \label{sectionV}}

\begin{figure}[h]
\centering
	\includegraphics[width=10cm]{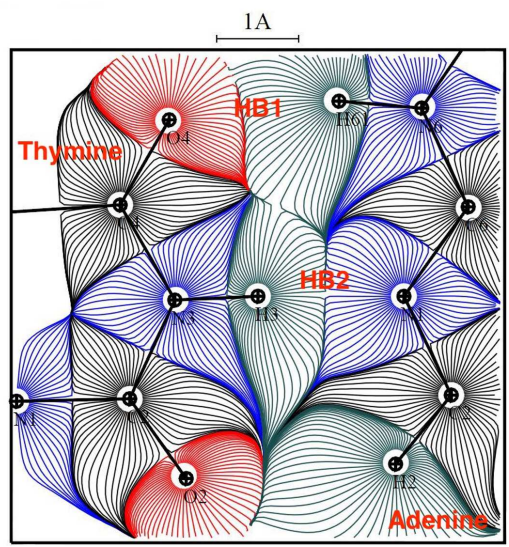}
	\caption{2D plot of the electrostatic potential gradient lines, in a plane containing the N3-H3...N1 Hydrogen bond. The figure depicts the electric potential atomic basins, delimited by surfaces of zero electric field flux, within which the total charge is zero. The lines are starting from nuclei and ending up where the electrostatic potential are a local minimum.}
	  \label{ElectricField}
\end{figure}

The previous estimate of the Rashba coupling contains a SO coupling strength $\xi_p$ and Stark contributions $\xi_{sp}^z$. While the SO coupling comes from the atomic strength (either N or O) coupling the onsite $p_z$ to $p_{x,y}$ orbitals, the Stark interaction depends on the extra electric field on the N and O due to the Hydrogen bond polarization. Here we analyze the Stark interaction produced on the oxygen and nitrogen sites in the immediate vicinity of a Hydrogen bond. Such sites possess a $\pi$ bond that carries mobile electrons that can be shared with vicinal base pairs. The Stark interaction is governed by $\langle 2p_x|e E(r)x|2s\rangle$ matrix element on the N or O sites. To estimate the electric fields $E(r)$, it is important to model the Hydrogen-bond beyond the dipole approximation which only captures the far field\cite{Blanco-Ruiz}. Figure \ref{ElectricField} shows the electrostatic potential gradient lines in the plane of the Adenine-Thymine base pair.
The field lines emanate away from the atomic cores starting from a minimum radius of $0$.$07$ \AA. The electric field is gradually shielded by the electron clouds both by their atoms own electronic density and the one either withdrawn or added due to electronegativity contrast. This shielding changes the electric field seen by valence electrons. In the figure we also see the surfaces of zero electric field flux (Bader surface) within which the total charge is null. On the limit between H3 and N1 basins lies a saddle point where the electrostatic potential is locally a minimum along the line, and maximum in directions perpendicular to it. Note the difference between the field lines at covalent bonds e.g. where they rapidly decrease in intensity as they approach the zero flux surface, and those of the two Hydrogen bonds that remain almost parallel until hitting the saddle point.  
\begin{figure}[h]
\centering
	\includegraphics[width=10cm]{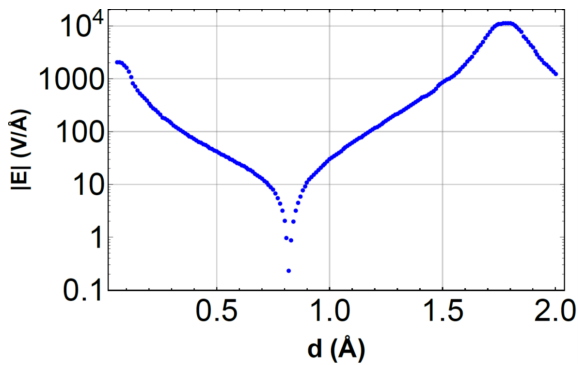}
	\caption{Modulus of the Electric field along Hydrogen bond HB2 (see Figure 2) starting from corresponding H atom. The dip in the electric field corresponds to the zero flux frontier between the hydrogen and the nitrogen partner (Bader surface). The derived profile is consistent with DFT calculations in ref.\cite{Blanco-Ruiz}.}
	  \label{FieldProfile}
\end{figure}
The electric field profile (along the hydrogen bond HB2) is depicted in Fig.\ref{FieldProfile}. There we can see the electric field intensities running from the positively polarized hydrogen to the core of the partnering nitrogen (see Appendix B). There is a sharp dip at the zero flux interface about which fields above 10 V/\AA ~arise very rapidly within a range of 0.25\AA. 

Using the computed electric fields we derived (see Appendix A) the Stark matrix elements arriving at parameter values depending on the particular HB between $\xi_{sp}^{x}\sim 10-20$ eV. The radial component of the field centered at N or O does not contribute to the integral by symmetry, so the contributing component is along the hydrogen bond axis ($x$ direction in the system of coordinates) within the range where both the $2s$ and $2p_x$ orbitals possess an appreciable density. Clearly the magnitude of the Stark coupling is at the limit of validity of perturbation theory since it is of the same order of magnitude as the energy levels themselves, due to the strength of the Hydrogen bond electric field.

\begin{figure}[h]
\centering
	\includegraphics[width=10cm]{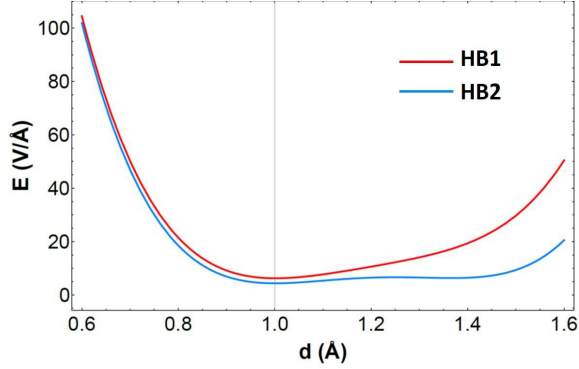}
	\caption{Electric field E in V/\r{A} versus the length $d_0$ of HB1 and HB2 at A/T bases. Plot based on values taken from Ref.\cite{Blanco-Ruiz}.}
	  \label{Evsd}
\end{figure}

As an additional check for the previous estimate we used a completely different source calculation for the electric fields beyond the point dipole model. Ruiz-Blanco et al\cite{Blanco-Ruiz} report that the electric field magnitude using state of the art DFT. For a distance of $1$.$5$ \r{A} it is of $35$ V/\r{A} and $15$ V/\r{A} for the Hydrogen bonds of the A-T bases. These values are really a lower bound for the electric fields since it is well established that the HB1 and HB2 distances can exceed $1.8$\r{A}\cite{PCCP,Wu,Fonseca}. An additional contribution of the previous work is that it estimates what happens with the Hydrogen bond axis electric field when the bond is stretched or contracted. We will use this study as a guide to understand what happens to the Stark coupling if the molecule is deformed. This way we can probe the model by looking at the resulting spin-polarization in the AFM mode\cite{NaamanSingleMolec}. Figure \ref{Evsd}, adapted from ref.\cite{Blanco-Ruiz}, reports the electric field as the Hydrogen bond is stretched, for both the HB1 and HB2. A mechanical model that accounts for the change in the Rashba coupling with changes in the hydrogen bond strength is proposed in the next section.

The main result of this paper is then that our model predicts a strong Rashba SO coupling (units to tens of meV) due to the strong electric field strength resulting from Hydrogen bonding bearing on $\pi$ bonds that couple bases for electronic transport. We believe these are the largest SO couplings one expects in organic molecules that exhibit CISS. Previously, detailed analytical tight binding calculations with either reasonable external electric fields and intrinsic SO couplings (purely atomic sources for the coupling) are a thousand and ten times smaller respectively\cite{Varela}. For empirical tight-binding models that fit the needed SO coupling to the observed polarization of filtered electrons the estimates are much larger\cite{GutierrezAdame}.

\section{Deformations as an orbital probe}
In this section we will derive the effects of changing the Hydrogen-bond polarization, on the Rashba spin-orbit strength predicted by our model. We then propose experiments that deform the DNA-helix model in characteristic ways to reveal the underlying physical process. We consider stretching and/or compressing the DNA model assuming two schemes\cite{Varela2018}: i) both strands of the DNA are pulled on each end of the segment\cite{Gupta}, and ii) one strand pulled on one end while the other is pulled on the other end in the opposite direction\cite{Kiran}. Both these deformation strategies assume that one or both strands are held. Nevertheless, scheme ii) can also be implemented as in ref.\cite{Paik} where the two strands on one end are allowed to rotate. The motivation for this study is that depending on the deformation scheme the Hydrogen-bond polarization will be differentially affected and therefore we will get an experimental signature effect on the electron polarization.
\begin{figure}
\centering
	\includegraphics[width=10cm]{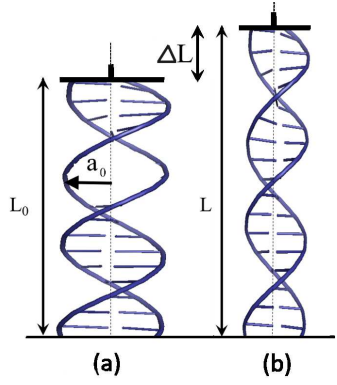}
	\caption{Deformation scheme I: Applied to a DNA segment of initial length $L_0$ and radius $a_0$. Panel (a) shows the unstretched situation of length $L_0$ and (b) stretching a length $\Delta L$ to a total length $L$  Both strands on each end are fixed and the double helix is mechanically stretched by $\Delta L$. The lateral width changes according to the Poisson ratio $\nu$ reported experimentally.}
	  \label{schemeI}
\end{figure}

The first deformation arrangement is shown in figure \ref{schemeI}. Geometrically this implies that the orbitals on the bases do not change their relative orientation and the angle turn per base $\Delta\phi$ remains unchanged for small deformations. The elastic behavior of the chain can be described by its Poisson ratio $\nu$. For double stranded DNA, it has been estimated from experiments\cite{Gupta} that $\nu=0.5$. The strain is defined as $\varepsilon=(L-L_0)/L_0=\Delta L/L_0$, where $L_0$ and $L$ are the initial and final lengths of the double helix, respectively. In this scheme, a change $\varepsilon$ in the length $L_0$, such that the final length is $L=L_0(1+\varepsilon)$, implies a change in the pitch  $b=b_0+(L_0/\mathcal{N})\varepsilon$ and in the radius $a=a_0(1-\nu\varepsilon)$ of the double helix, where $b_0$ and $a_0$ are the corresponding parameters without deformation and $\mathcal{N}$ is the number of turns of the helix. These changes modify the distance between consecutive bases on each strand and most importantly the length of the hydrogen bond that connects the bases. We can impose these constraints on the expression for the Rashba SO coupling since we have explicit dependences on all geometrical variables. The expression for $\lambda_R$ in this scheme is 
\begin{widetext}
\begin{equation}
    \lambda_{R}(\varepsilon)=\frac{8\pi^2\hbar^2\kappa_{ps}\xi_p [\xi_{sp}^{(\imath)}(\varepsilon)+\xi_{sp}^{(\jmath)}(\varepsilon)]\left(b_0+\frac{L_0}{\mathcal{N}}\varepsilon\right)\Delta\phi}{(\epsilon_{2p}^{\pi}-\epsilon_s)(\epsilon_{2p}^{\pi}-\epsilon_{2p}^{\sigma})m\left[\left(b_0+\frac{L_0}{\mathcal{N}}\varepsilon\right)^2\Delta\phi^2 - 8\pi^2a_0^2(1-\nu\varepsilon)^2(\cos{\phi}-1)\right]^{3/2}}.
\end{equation}
\end{widetext}
The Stark parameters will be modulated by the change in the hydrogen bond polarization due to the deformation. We consider the polarization change from the results of ref.\cite{Blanco-Ruiz}.  
Stretching the chain increases the distance between the bases $R_{\imath\jmath}$ and therefore decreases the orbital overlap $E_{sz}^{\imath\jmath}$, whereas on compression within an acceptable deformation range, $R_{\imath\jmath}$ decreases and $E_{sx}^{\imath\jmath}$ is enhanced. Furthermore, stretching decreases the radius of the helix. Such a decrease is assumed to be absorbed by the distance $d_0$ of the HB1/HB2 causing a concomitant decrease in the polarity of the hydrogen bond below the value at zero deformation. 

\begin{figure}
	\includegraphics[width=10cm]{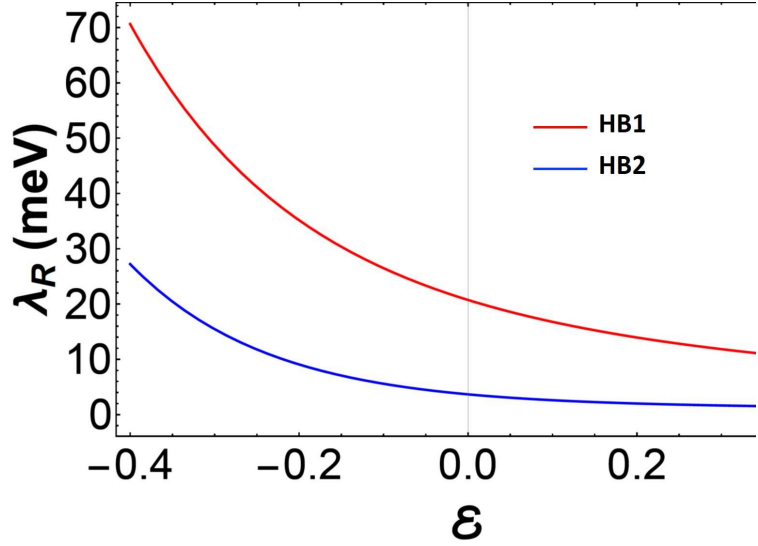}
	\caption{Rashba magnitude $\lambda_R$ versus deformation $\varepsilon$ under scheme I setup. We used $d_0=2a_0=1.74$\r{A}, $b_0=35.4$\r{A} and $\Delta\phi=\pi/5$. For $\varepsilon=0$, intensity of the interaction are $\sim 3.6$ meV and $\sim 20$ meV for HB2 and HB1, respectively. Stretching the helix ($\varepsilon>0$) decreases the Rashba coupling while compressing increases it.}
	  \label{Lambda1}
\end{figure}

Since $\lambda_R$ is proportional to electric field and orbital overlaps,  $\lambda_R$ will follow the same behaviors as these parameters under deformations as shown in the figure \ref{Lambda1} in a deformation regime of up to $40$\%. Note that the Rashba coupling decreases under stretching since the hydrogen bond polarization is decreased by the ensuing compression due to the reduction of the radius of the molecule. The opposite behavior is seen on compressing the molecule with a steeper slope.

The hydrogen bonding for an Oligopeptide, such as the one studied in Kiran et al\cite{Kiran}, has a very different disposition, joining different turns of the helix. In an Oligopeptide the polarization of the HB is increased when the molecule is stretched while in scheme I the HB is expected to decrease its polarization for a small deformations. This scenario is clearly consistent with the experimental results of ref.\cite{Kiran}. 

\begin{figure}[h]
	\includegraphics[width=10cm]{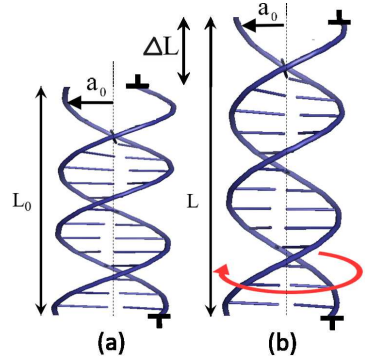}
	\caption{Deformation scheme II: One strand fixed on one end the opposite strand fixed on the other end. Panel (a) shows the unstretched situation of length $L_0$ and (b) stretching a length $\Delta L$ to a total length $L$ . In this setup the molecule rotate on stretching changing minimally the Hydrogen bond polarization. The changes in the SO coupling then only depend on orbital overlaps.}
	  \label{SchemeII}
\end{figure}

For the second deformation setup (see Fig.\ref{SchemeII}), we load the double helix so that one strand is pulled on one end and the opposite strand on the other end. The radius of the helix and the distance between consecutive bases is kept constant while the angle $\Delta\phi$ and the pitch $b$ change together. The relation between pitch and rotation per base is 
\begin{equation}
    b=\frac{2\pi}{\Delta\phi }R_{\imath\jmath}\sqrt{1-4\left(\frac{a_0}{R_{\imath\jmath}}\right)^2 \sin^2(\Delta\phi/2)},
\end{equation}
where the overlap $E_{sp}^{\imath\jmath}$ changes with deformation in the form
\begin{equation}
E_{sz}^{\imath\jmath}(\varepsilon)=V_{sp}^{\sigma}\sqrt{1-4\left(\frac{a_0}{R_{ij}}\right)^2\sin^2\left(\frac{\Delta\phi(1-\varepsilon)}{2}\right)}. 
\label{EszSchemeII}
\end{equation}

In this deformation scheme the radius remains constant, the polarization of the hydrogen bond does not change while the orbital overlaps do, such that the magnitude of the Rashba interaction results in
\begin{widetext}
\begin{equation}
    \lambda_R(\varepsilon)= \left(\frac{\kappa_{ps}\hbar^2\xi_p(\xi_{sp}^{(\imath)}+\xi_{sp}^{(\jmath)})}{mR_{\imath\jmath}^2(\epsilon_{2p}^{\pi}-\epsilon_s)(\epsilon_{2p}^{\pi}-\epsilon_{2p}^{\sigma})}\right)\sqrt{1-4\left(\frac{a_0}{R_{ij}}\right)^2\sin^2\left(\frac{\Delta\phi(1-\varepsilon)}{2}\right)},
\end{equation}
\end{widetext}
where the first term in expression remains invariant with deformation. 
\begin{figure}[h]
\centering
	\includegraphics[width=10cm]{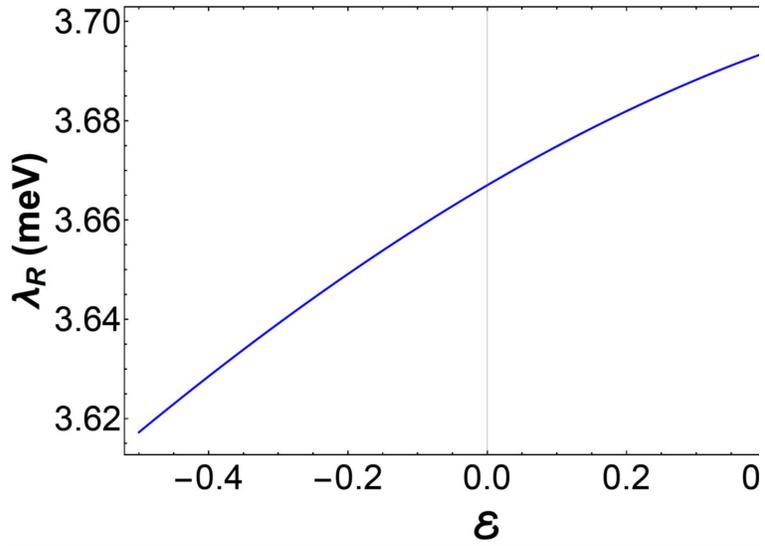}
	\caption{Rashba magnitude $\lambda_R$ of HB1 versus deformation $\varepsilon$ for the scheme II setup. We used $b_0=35.4$ \r{A}, $\Delta\phi=\pi/5$, and $a_0=1.74$ \r{A}.}
	  \label{Lambda3}
\end{figure}
The behavior of $\lambda_R$ as a function of $\varepsilon$ is shown in figure \ref{Lambda3}. A stretching implies a decrease of $\Delta\phi$ and by equation (\ref{EszSchemeII}), $E_{sz}^{\imath\jmath}$ increases, which induces an increase in the magnitude of $\lambda_R$. The variation of the magnitude in this case, compared to the previous scheme, is small of approximately $0.2$ \% for a wide range of $\varepsilon$. This indicates that according to our model, the strength of the Rashba interaction is strongly dependent on the polarization associated  with the inter-base Hydrogen bonds in DNA. 

\section{Summary and conclusions}
We have derived a simple model for the effect of the Hydrogen bond polarization in giving rise to an intrinsic Rashba coupling in DNA. The basic ingredients of the model contemplate transport electrons associated with $\pi$ orbitals on the DNA bases on atoms that are part of a Hydrogen bond. By band folding perturbation theory we assess that the spin activity of the molecule (Rashba coupling) is governed by the combined effect of the intrinsic SO coupling of the double bonded atom on the base (O and N) and the Stark interaction due to the hydrogen bond polarization. The intrinsic Rashba effect (all internal fields of the molecule) is predicted to be the largest yet found from a detailed model and should be the source of any spin activity with the same symmetry properties of the CISS effect. We find a Rashba term from 3.6-20 meV depending on the particular HB, coming from a considerable Stark coupling in the range of 10-20 eV due to the hydrogen bond polarization. We propose specific experimental setups to prove the details of the model directly through a mechanical spectroscopy of sorts, that exposes the role of the Hydrogen bonding in the molecular spin activity. Although more precise calculations are necessary assessing details of the molecular structure potentially involved, we believe that our estimate is a realistic order of magnitude estimate for DNA. Although there are other sources of SO coupling, the Hydrogen bond source is one order of magnitude larger than purely atomic SO couplings and three to six order of magnitudes larger than extrinsic Rashba terms. We also point out that the highly polarized Hydrogen bond is also present in oligopeptides (responsible for stabilizing helical structure) that also display a strong spin activity \cite{Aragones}.  We expect similar results for these structures and speculate that Hydrogen bonding is responsible for the strongest spin activity seen for CISS effects in biological chiral molecules.

\begin{acknowledgments}
This work was supported
by grant ``CEPRA XII-2108-06 Espectroscop\'ia
Mec\'anica'' of CEDIA, Ecuador and exchanges between Ecuador and France were supported by the PICS CNRS "Scratch it" project. 
\end{acknowledgments}

\appendix

\section{The magnitude of the Stark Interaction }

The Hamiltonian that represents the Stark interaction is given by $H_s=-e\vec{E}\cdot\vec{r}$, where $\vec{E}$ is an external electric field that induces the interaction, and $\vec{r}=x\hat{\imath}+y\hat{\jmath}+z\hat{k}$. In presence of an electric field $\vec{E}=E_x\hat{\imath}+E_y\hat{\jmath}+E_z\hat{k}$, $H_s$ can be written in spherical coordinates as
\begin{equation}
    H_s=-eE_xr\sin{\theta}\cos{\varphi}-eE_yr\sin{\theta}\sin{\varphi}-eE_zr\cos{\theta},
\end{equation}
$e$ being the electric charge.

\begin{figure}
\centering
	\includegraphics[ width=10cm]{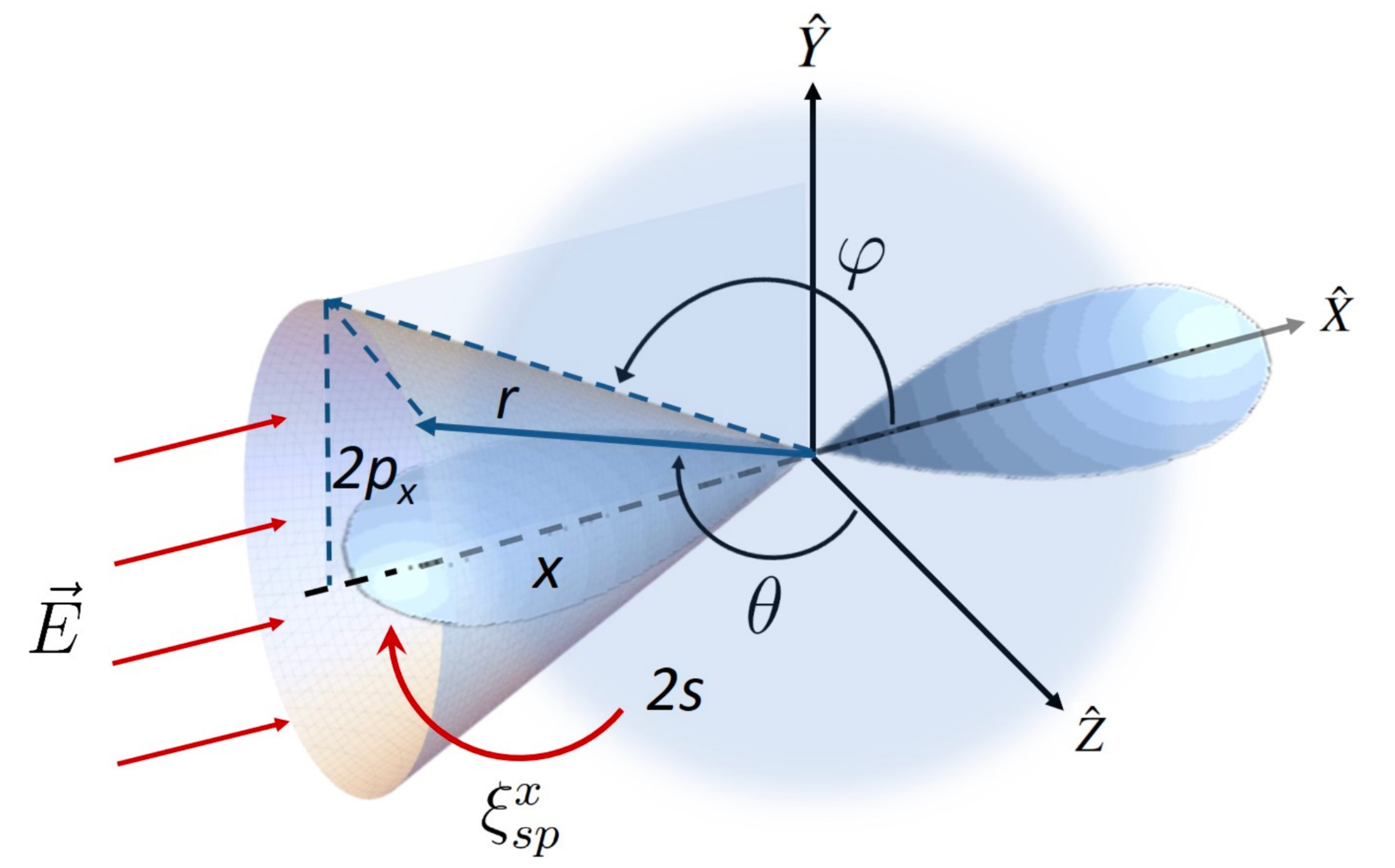}
	\caption{Figure depicts $s$ and $p_x$ orbitals and the square cone region of integration.}
	  \label{Stark1}
\end{figure}

The Stark interaction couples hydrogenic orbitals of the atom in the form
\begin{equation}
\xi_{sp}^i=\bra{s}H_s\ket{p_i}=\int_0^{2\pi}\int_0^{\pi}\int_{0}^{\infty}s^*H_s p_i r^2 \sin{\theta}dr d\theta d\varphi,
\label{Stark3}
\end{equation}
where $i=x,y,z$ representing the $p$ orbitals. In our case, we consider that the electric field of hydrogen bond permeates the Nitrogen or Oxygen $p_x$ orbital aligned with the hydrogen bond. The electric field is greatest along the hydrogen bond axis and $E_x$ decays quickly off axis (See Fig. 3 of article). The integration zone is given by angular range along of axis $x$, in the negative direction, where the electric field is assumed as $\vec{E}=\vec{E}(x)$ and therefore the only non-zero element for the interaction is $\xi_{sp}^x=\bra{2s}H_s\ket{2p_x}$.  (see Fig.\ref{Stark1}). 

\begin{figure}[h]
\centering
	\includegraphics[ width=10cm]{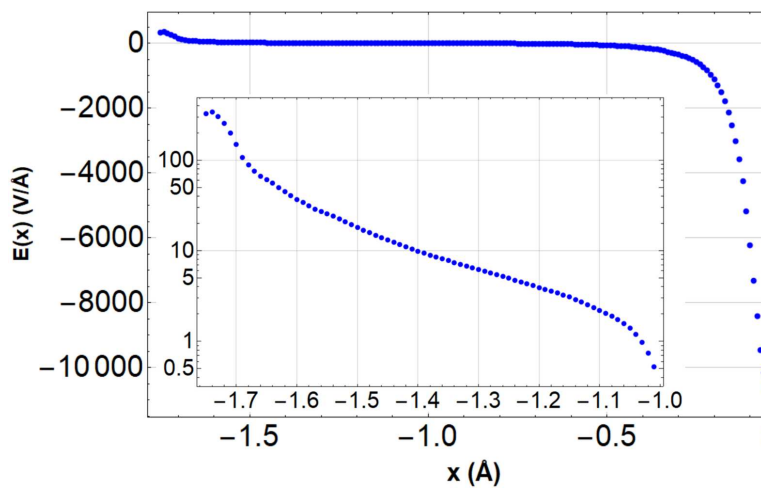}
	\caption{Electric field along the axis of the hydrogen bond. The coordinate $x=0$ corresponds to the nitrogen center. The main panel shows the full range of the hydrogen bond, and the inset shows the behavior of the electric field in the region furthest from the nitrogen nucleus.}
	  \label{FieldEx}
\end{figure}

For an atom with atomic number Z, the hydrogenic orbitals $2s$ and $2p_x$ in spherical coordinates are: 
\begin{eqnarray}
\ket{2s}&=&\sqrt{\frac{Z^3}{8\pi a_0^3}}e^{-Zr/2a_0}\left(1-\frac{Zr}{2a_0}\right)\nonumber\\
\ket{2p_x}&=& \sqrt{\frac{Z^5}{32\pi a_0^3}}e^{-Zr/2a_0}\left(\frac{Zr}{a_0}\right)\sin{\theta}\cos{\varphi},
\end{eqnarray}
where $a_0$ is Bohr's radius. Then, 
\begin{widetext}
\begin{eqnarray}
\xi_{sp}^x=-\frac{Z^5e}{16a_0^4\pi\sqrt{2}}\int_{\varphi_1}^{\varphi_2}\int_{\theta_1}^{\theta_2}\int_{r'}E_x x  \left(1-\frac{Zr}{2a_0}\right)e^{-Zr/a_0}r^3 \sin^2{(\theta)}\cos{(\varphi)}dr d\theta d\varphi,
\label{starkx1}
\end{eqnarray}
\end{widetext}

and $r'$ indicates that integral is in the spacial coordinate. 
The electric field profile along the hydrogen bond to HB2 is shown in figure \ref{FieldEx}.

To solve integral \ref{starkx1}, we consider a change of variables such that for a fixed value of $x$ coordinate, $r$ varies with $\theta$ and $\varphi$ in the form (see figure \ref{Stark1})
\begin{equation}
    r=\frac{x}{\sin{\theta}\cos{\varphi}}. 
\end{equation}
The matrix element $\xi_{sp}^x$ is finally given by
\begin{widetext}
\begin{equation}
    \xi_{sp}^x(x,\theta,\varphi)=-\frac{Z^5e}{16a_0^2 \pi \sqrt{2}}\int_{\varphi_1}^{\varphi_2}\int_{\theta_1}^{\theta_2}\int_{x_1}^{x_2}E_x  \left(1-\frac{Zx}{2a_0\sin{\theta}\cos{\varphi}}\right)e^{-\frac{Zx}{a_0\sin{\theta}\cos{\varphi}}}\left(\frac{x^4}{\sin^2{\theta}\cos^3{\varphi}}\right)dx d\theta d\varphi
\label{Stark2}
\end{equation}
\end{widetext}

The electric field shown in figure \ref{FieldEx} contains the contribution of the atom on site. As the radial electric field gives a null Stark matrix element coupling for the full angular range we need to subtract the electric field contribution from the atom. For a hydrogenic atom for Nitrogen, considering spherical symmetry, the charge density is
\begin{equation}
    \rho(r)=\frac{e}{\pi}\left(\frac{Z}{a_0}\right)^3 e^{-2Zr/a_0}. 
\end{equation}
Using Gauss' Law, the electric field is
\begin{equation}
\vec{E}(r)=\frac{e}{4\pi\epsilon_0 r^2}\left[1-e^{-2Zr/a_0}\left(1+\frac{2Zr}{a_0}+2\left(\frac{Zr}{a_0}\right)^2\right) \right]\hat{r},
\end{equation}
with $\epsilon_0$ the electric permittivity in the vacuum. 

In the region of interest the value of the electric field is of the order of $800$ V/\r{A}. The Stark term associated with this field is given by equation \ref{Stark3} on appropriated limits.

Finally, the effective coupling Rashba is 
\begin{equation}
    \lambda_R^x=-\frac{2 E_{sz}^{\imath\jmath}\xi_p\xi_{sp}^x}{(\epsilon_{2p}^{\pi}-\epsilon_s)(\epsilon_{2p}^{\pi}-\epsilon_{2p}^{\sigma})}.
\end{equation}

Results of $\xi_{sp}^x$ for $x\in [0,1.8]$ (Hydrogen bond length) and differents values of  $\theta\in [\theta_1,\theta_2]$ and $\varphi\in [\varphi_1,\varphi_2]$, and $\lambda_R^x$ for Nitrogen is shown in Table \ref{Table1}.

\begin{center}
\begin{table}[h]
\begin{tabular}{|c|c|c|c|}
\hline 
 $\theta\left[\theta_1,\theta_2\right]$& $\varphi\left[\varphi_1,\varphi_2\right]$ & $\xi_{sp}^x$ (eV) & $\lambda_R^x$ (eV)\\ 
\hline 
$\left[\frac{89\pi}{180},\frac{91\pi}{180}\right]$ & $\left[\frac{179\pi}{180},\frac{181\pi}{180}\right]$ & -0.0486 & $1.37\times 10^{-5}$ \\ 
\hline 
$\left[\frac{17\pi}{36},\frac{19\pi}{36}\right]$ & $\left[\frac{35\pi}{180},\frac{37\pi}{180}\right]$ & -1.21799 & $3.50\times 10^{-4}$\\ 
\hline 
$\left[\frac{5\pi}{12},\frac{7\pi}{12}\right]$ & $\left[\frac{11\pi}{12},\frac{13\pi}{12}\right]$ & -11.1911 & $3.67\times 10^{-3}$\\ 
\hline 
\end{tabular} 
\caption{Convergence to values for the Stark matrix element and the Rashba coupling as the full effect of the electric field due to the hydrogen bond polarization is included in the integration range.}
\label{Table1}
\end{table}
\end{center}
The $x$ component of the electric field is essentially zero beyond the third range of angles in the Table, so the estimated value for the Stark matrix element and the Rashba coupling is $~11.2$ eV and $3.7$ meV respectively, as reported in the article.


\section{Electric field computations}

Electric field magnitude values were computed on the experimental crystal structure of a DNA dodecamer (D(CGCGAATTCGCG)) deposited in the Protein Data Bank under the code 6CQ3. Hydrogen atoms were generated at the geometries expected from neutron diffraction using the MolProbity web server \cite{MolProbity}. In other words, hydrogen atoms nuclei are located at their "true" position, instead of at the position of their valence electron as would have indicated X-ray diffraction data. An explicit model of molecular electron density was then reconstructed for the 6CQ3 DNA crystal structure using atomic parameters transferred from the ELMAM2 library \cite{Elmam}. These parameters, written in the Hansen and Coppens multipolar formalism, describe atomic electron densities of biological atom types as sums of weighted nuclei-centered real spherical harmonics functions \cite{Coppens} reproducing the deformation of atomic electron clouds upon covalent bond formation.

The atomic parameters described in the ELMAM2 library used in this study, are of experimental origin. They are issued from averages of parameters obtained after refinement of small-molecules charge densities against subatomic resolution X-ray diffraction data.

After transfer of the electron density parameters from the ELMAM2 library, the electric field vectors were computed in two steps. First, using the MoPro \cite{MoPro} software the total electrostatic potential in the vicinity of the Thymine 8 - Adenine 17 N-H...N hydrogen bond (HB2 in the text) was analytically computed on a 1.7$\times$1.0$\times$1.0\r{A} regular rectangular grid with 0.005\r{A} sampling in each (orthogonal) direction, accounting for contributions of every atoms in the DNA dodecamer structure. Next, the electric field vectors were obtained on points of a similar sized and sampled grid through numerical differentiation of the electrostatic potential using a sixth order Taylor expansion formula. The values, initially computed in e/\r{A}, were finally scaled to the correct unit system (GV/m), assuming an \textit{in vacuo} dielectric constant.

\bibliography{biblio.bib}

\end{document}